\documentstyle[aps,preprint,epsfig,epsf,side,rotating,fancyheadings]{revtex}
\newfont{\trpolice}{cmbxti10 scaled \magstep1 }
\newfont{\trdpolice}{cmbxti10 scaled \magstep2 }
\newfont{\trrpolice}{cmti10 scaled \magstep1 }
\newfont{\trepolice}{cmti12 scaled \magstep1 }
\newfont{\treepolice}{cmti12 scaled \magstep2 }
\newfont{\treeepolice}{cmti12 scaled \magstep3 }
\begin {document}
\draft
\title{\Large \bf Spin-Alignment and Quasi-Molecular Resonance
 in Heavy-Ion Collision}
\author {R. Nouicer, C. Beck, N. Aissaoui, T. Bellot, G. de France, D.
Disdier, G. Duch\`ene, A.~Elanique, R.M.~Freeman, F. Haas, A. Hachem, 
F. Hoellinger, D. Mahboub, V. Rauch }
 
\address{\it Institut de Recherches Subatomiques, IN2P3-CNRS/Universit\'e Louis
Pasteur,\\
B.P.28, F-67037 Strasbourg Cedex 2, France }
 
\author { S.J. Sanders, T. Catterson, A. Dummer, F.W. Prosser }
 
\address {\it Department of Physics and Astronomy, University of Kansas,
Lawrence, USA }
 
\author { A. Szanto de Toledo }
 
\address {\it Depatamento de Fisica Nuclear, University of Sa\~o Paulo, Brazil }
 
\author { Sl. Cavallaro}
 
\address {\it Dipartimento di Fisica dell'Universit\'a di Catania, INFN
Catania, I-95129 Catania, Italy }
 
\author { E. Uegaki}
 
\address {\it Department of Physics, Akita University, Akita 010, Japan}
 
\author { Y. Abe }
 
\address {\it Yukawa Institute for Theoretical Physics, Kyoto University,
Kyoto 606, Japan }
 
\maketitle
\vskip 3cm
\centerline{\it \bf Invited talk presented at the 36$^{th}$ International
Winter Meeting on {\it ''Nuclear Physics''}}
\centerline{\it {\bf January 26-31, 1998} Bormio, Italy}

\newpage
 
\begin{abstract}
{\sl
Fragment-fragment-$\gamma$ triple coincidence measurements of the
$^{28}$Si~$+$~$^{28}$Si reaction at E$_{\rm c.m.}$~=~55.8~MeV, carefully chosen
to populate a well known quasi-molecular resonance in $^{56}$Ni, have been
performed at the VIVITRON tandem facility by using the Eurogam Phase~II
$\gamma$-ray spectrometer. In the $^{28}$Si~$+$~$^{28}$Si\ reaction, the
resonant behavior of the $^{28}$Si $+$ $^{28}$Si exit-channel is clearly
observed by the present fragment-fragment coincidence data. The more unexpected
result is the spin disalignment of the $^{28}$Si $+$ $^{28}$Si resonance. This
has been demonstrated first by the measured angular distributions of the
elastic 0$^{+}$, inelastic 2$^{+}$ and mutual excitation channels
2$^{+}$-2$^{+}$, which are dominated by a unique and pure partial wave with L =
38\ ${\rm \hbar}$, and has been confirmed by measuring their particle-$\gamma$
angular correlations with Eurogam Phase II. The spin disalignment supports new
molecular model predictions, in which the observed resonance would correspond
to the ''Butterfly mode''. A discussion concerning the {\it spin alignment and
spin disalignment} for different systems : $^{12}$C~$+$~$^{12}$C,
$^{24}$Mg~$+$~$^{24}$Mg and $^{28}$Si~$+$~$^{28}$Si will be given. } 
\end{abstract}

\vskip 2.0cm
 
{ PACS number(s): 25.70.Ef, 25.70.Jj, 25.70.Gh, 25.70.Lm,24.60.Dr, 23.20.Lv }

\newpage

\centerline{\Large \sl \bf I. Introduction }
\vskip 0.7cm
One of the most intriguing results in the study of heavy-ion resonances has
been the observation of the pronounced, narrow and well isolated
quasi-molecular resonances in the elastic and inelastic scattering of medium
mass nuclei. The first observation of such resonant structures in systems
around the mass A = 50 region was reported by Betts et al. [1] in $^{28}$Si~$+$~$^{28}$Si. 
Even more pronounced resonance structures were later reported in
$^{24}$Mg~$+$~$^{24}$Mg [2] system, while other symmetric systems such as
$^{32}$S~$+$~$^{32}$S [3] and $^{40}$Ca $+$ $^{40}$Ca [6] do not show such
behavior. Resonance structures were also found in the asymmetric system
$^{24}$Mg~$+$~$^{28}$Si [4]. The appearance of narrow resonances in
medium-heavy systems may open up the possibillity for studying nuclear
structure at high spin in the continuum of the complex dinuclear system. The
excitation functions for elastic and inelastic scattering for several low-lying
excitations in $^{28}$Si $+$ $^{28}$Si and $^{24}$Mg $+$ $^{24}$Mg are shown in
Fig.1 (figures taken from [1,2]). Perhaps the most intriguing aspect
of these excitation functions is the narrow width structures which are observed
to be very highly correlated. 
\par
For a better understanding of the nature of these resonances it is important to
know their detailed properties. Most important is the determination of their
spins. One of the explanations suggested for the nature of these resonances is
that they are associated with quasi-stable configurations with extreme
deformation. This interpretation is supported by Nilsson-Strutinsky model
calculations. In this model the structure of the compound nucleus at large
deformation and high angular momenta has been considered. Calculations of the 
potential energy surfaces for nuclei such as $^{56}$Ni and $^{48}$Cr using the
Nilsson-Strutinsky prescription show significant secondary minima
($\beta$~=~0.9) at extremely large prolate deformations. An example of such a
potential energy surface is shown in Fig.2 (figure taken from [5,6]),
for the nuclei $^{56}$Ni at spin I = 40 and $^{48}$Cr at spin I~=~36
respectively. 
\par
In this paper, we present experimental results for the $^{28}$Si~$+$~$^{28}$Si
reaction using very powerful coincidence (fragment-fragment-gamma : Eurogam
Phase II with ancillary detectors) techniques. These results supply, for the
first time in a heavy-ion resonance, evidence for a spin disalignment between
the binary fragments spins and the total angular momentum. This result which
will be interpreted in the framework of a molecular model by a ''Butterfly
motion''. The search for highly deformed bands in the $^{28}$Si nucleus
produced in the $^{28}$Si~$+$~$^{28}$Si reaction at a resonance energy E$_{\rm
lab}$~=~111.6~MeV is presented. A discussion concerning the {\it spin alignment
and disalignment} for the following  systems $^{12}$C~$+$~$^{12}$C,
$^{24}$Mg~$+$~$^{24}$Mg and $^{28}$Si~$+$~$^{28}$Si will be given. 
\vskip 0.7cm
\centerline{\Large \sl \bf II. Experimental Techniques } 
\centerline{\large \sl ''Eurogam Phase II with Ancillary Detectors'' }
\vskip 0.7cm
The experiment $^{28}$Si + $^{28}$Si was performed at the IReS Strasbourg
VIVITRON tandem facility with a $^{28}$Si beam at the bombarding energy E$_{\rm
lab}$ ($^{28}$Si) = 111.6\ MeV, carefully chosen to populate the well known
38$^{+}$ resonance [7]. The $^{28}$Si beam was used to bombard a 25 ${\rm \mu g
/cm^{2}}$ thick $^{28}$Si target. The thickness of the target corresponds to an
energy loss of 130 keV, which is smaller than the resonance width ($\Gamma_{\rm
c.m.} \approx $ 150\ keV). The experiment has been carried out in a triple
coincidence mode (fragment-fragment-$\gamma$). The fission fragments were
detected in two pairs of large-area position-sensitive Si(surface-barrier)
detectors (PSD) placed on either side of the beam axis and their masses were
determinated by using standard kinematic coincidence techniques [8]. The
opening angle of the PSD's covered in the laboratory the angular range 22$^{o}$
to 73$^{o}$ in the horizontal plane. The $\gamma$-rays were detected in the
Eurogam Phase II multi-detector array [9], which consists of 54
Compton-suppressed germanium (Ge) detectors, 30 tapered coaxial Ge detectors
from Eurogam Phase I located in the forward and backward hemispheres, and 24
clover detectors installed at~$\approx$~90$^{o}$ relative to the beam axis. The
number of the Ge crystals at each angle with respect to the beam direction is
(5, 22$^{o}$), (10, 46$^{o}$), (24, 71$^{o}$), (24, 80$^{o}$), (24, 100$^{o}$),
(24, 109$^{o}$) (10, 134$^{o}$) and (5,~158$^{o}$). Energy and relative
efficiency calibrations of Eurogam Phase II were obtained with standard
$\gamma$-ray sources and an AmBe source for the higher energy $\gamma$-ray
region [9,10,11]. The detectors associated with the experiment
$^{28}$Si~$+$~$^{28}$Si (fragment-fragment-$\gamma$) are shown in
Fig.3. The $\gamma$-rays emitted by excited fragments are detected by
the Ge detectors of Eurogam Phase II. Doppler-shift corrections were applied to
the $\gamma$-ray data on an event-by-event basis using measured velocities and
angles of the detected fragments. 
\vskip 0.7cm
\centerline{\Large \sl \bf III. Experimental Results of $^{28}$Si $+$ $^{28}$Si 
Reaction}
\vskip 0.7cm
Fig.4 shows the two-dimensional energy spectrum E$_{3}$-E$_{4}$ of
the fragments in coincidence with one pair of position-sensitive detectors.
Three regions, noted 1, 2 and 3, can be distinguished in the figure. The region
1 corresponds to the binary products of the $^{28}$Si~$+$~$^{28}$Si reaction.
In this region the spectrum shows well structured distributions indicating that
the bombarding energy corresponds well to the resonance energy. The region~2
arises from the reaction products of the $^{28}$Si with heavy contaminants, and
region~3 corresponds to the light charged particle ($p$, $\alpha$) coincidences
arising from the fusion-evaporation of the $^{56}$Ni compound system. The
identification of different exit channels corresponding to region~1 has been
obtained by using mass spectra (as shown in Fig.5) constructed by
standard binary kinematic relations [9]. 
\par
Fig.5 shows that the region 1 is formed by two kinds of reaction. The
first (Fig.5.A) corresponds to reaction products of $^{28}$Si $+$
$^{28}$Si at resonance energy E$_{\rm lab}$~=~111.6\ MeV and shows that the
reaction is dominated by the $^{28}$Si $+$ $^{28}$Si exit-channel. This is due
to the fact that the PSD's locations have been optimised to study this
exit-channel. The second reaction (Fig.5.B) arises from the reaction
$^{28}$Si on the $^{16}$O target contaminant. 
\par
In order to understand better the nature of the quasi-molecular resonances it
is important to know their detailed {\it spin} properties. For this purpose we
will focus our analysis on the $^{28}$Si $+$ $^{28}$Si exit-channel. 
\newpage
\centerline{\Large \sl \bf A. Fragment-Fragment Coincidence Measurements}
\centerline{\large \sl ``Observation of Spin Disalignment ''}
\vskip 0.7cm
By setting gates with the conditions M$_{\rm T}$ = 56 and M$_{\rm 3}$ =28 in the
mass spectra, we have selected the $^{28}$Si~$+$~$^{28}$Si exit-channel [9].
This has permitted us to construct the excitation energy spectra (E$_{\rm X}$)
defined by the following expression : 
\begin{equation}
E_{X} = Q_{gg} - Q^{*} = Q_{gg} + E_{\rm lab} - (E_{3} +E_{4})
\end{equation}
\noindent where $Q_{gg}$ is the reaction $Q$-value. E$_{3}$ and E$_{4}$ are the
energies of the two fragments detected in coincidence. 
\par
Fig.6 shows the gated two-dimensional spectrum of the ejectile energy
E$_{\rm 3}$ as a function of the excitation energy E$_{\rm X}$ for the
$^{28}$Si~$+$~$^{28}$Si exit-channel. The vertical band corresponds to
different excited states in the two $^{28}$Si fragments, whereas the regular
pattern of yields are due to strongly structured angular distributions. In
order to investigate the resonance effects, we will extracted the angular
distributions from the projections on the E$_{\rm X}$ axis with excitation
energy gates around E$_{\rm X} \approx $ 0, 1.7 MeV, 4 MeV, 6 MeV and 10 MeV
corresponding to the different states of the two $^{28}$Si fragments. This
method, which allows us to extract the angular distributions at the different
excitation energies, will permit us to determine the dominant orbital angular
momentum characterzing the distributions and to distinguish between resonance
and fission processes. 
\par
The angular distributions (AD) for identical particle exit-channel
$^{28}$Si~$+$~$^{28}$Si at different excitation energies are displayed in
Fig.7. We can distinguish two trends in the AD behavior~: 
\vskip 0.4cm
\noindent{\trpolice 1 - At the low-lying states E$_{\rm X} \le $ 4 MeV :}
\vskip 0.4cm
The $^{28}$Si($^{28}$Si,$^{28}$Si)$^{28}$Si identical particle exit-channel at
E$_{\rm lab}$ = 111.6 MeV was found to have at backward angles (between 70$^{o}
\le \theta_{c.m.} \le $ 110$^{o}$) strongly oscillatory angular distributions
for the elastic, inelastic, and mutual excitation channels as shown in
Fig.7. The present large-angle high-quality data, with good
position resolution and high statistics, are well described by the curves of 
Fig.7 as calculated by [P$_{L}(\cos \theta_{c.m.} )$]$^{2}$ shapes
with L~=~38~${\rm \hbar}$ in perfect agreement in the elastic channel with the
previous data of Betts et al. [7]. The fact that the measured angular
distributions in the elastic, inelastic, and mutual excitation channels
correspond to shapes characterized by the same single Legendre polynomial
squared means that the resonant behavior is dominated by a unique and pure
partial wave associated with the angular momentum value L = 38 ${\rm \hbar}$.
This value can finally be considered as the spin of the well defined and
isolated quasi-molecular resonance. 
\par
{\it Knowing that the total angular momentum 
$\vec{J} = \vec{L} + \vec{I}$ is conserved and that the angular momentum 
L = 38 ${\rm \hbar}$ is dominant in these three resonant channels
implies that the projection of the spin along the direction perpendicular to
the reaction plane is {\it m =0}. This is the signature of the disalignment 
between the relative angular momentum $\vec{L}$ and total spin $\vec{I}$.} 
\par
This signature of spin disalignment will be verified in the following by
fragment-fragment-gamma correlation measurements for the two low-energy
inelastic excitations using the Eurogam Phase II multi-detector array. 
\vskip 0.4cm
\noindent{\trpolice 2 - At the high excitation states E$_{\rm X} \ge $ 6 MeV :}
\vskip 0.4cm
The identification of the excited states (3$_{1}^{-}$, 0$_{1}^{+}$),
(4$_{1}^{+}$, 2$_{1}^{+}$) and (4$_{1}^{+}$, 4$_{1}^{+}$) as shown in
Fig.7 at high excitation energy E$_{\rm X} \ge $ 6 MeV has been
obtained by using $\gamma$-ray coincident data. The AD for mutual excited
states (3$_{1}^{-}$, 0$_{1}^{+}$) and (4$_{1}^{+}$, 2$_{1}^{+}$) at the
excitation energy E$_{X}\approx$~6~MeV are not so strongly structured as
compared to the low-lying states and it is difficult to fit them by the
Legendre polynomial squared. Therefore the AD for mutual excited states
(4$_{1}^{+}$, 4$_{1}^{+}$) has a shape comparable to a 1/$\sin\theta_{c.m.}$
behavior and we can consider the two $^{28}$Si fragments as being emitted from
a fully relaxed process such as {\it fusion-fission}. 
\par
In general the angular distributions show that the low-lying states are
characterised by the resonance phenomena and, for the first time, by {\it spin
disalignment}. The fission process is believed to dominate the distribution at
high excitation energy E$_{\rm X}\ge$~6~MeV. 
\newpage
\centerline{\Large \sl \bf B. Fragment-Fragment-Gamma Coincidence Measurements}
\centerline{\large \sl ``Confirmation of Spin Disalignment ''}
\vskip 0.5cm
In order to understand this disalignment we will focus our analysis on the
fragment-fragment-$\gamma$ coincidence data of the $^{28}$Si + $^{28}$Si 
exit-channel [9,10,11]. Spin disalignment estimates of the low-lying excitation
states (single inelastic 2$^{+}_{1}$ and mutual inelastic (2$^{+}_{1}$,
2$^{+}_{1}$) exit-channels) have been deduced by measuring their
particle-$\gamma$ angular correlations with Eurogam Phase II. 
\par
In Fig.9 three quantization axes have been defined as follows : a)
corresponds to the beam axis, b) axis normal to the scattering plane, and c)
axis perpendicular to the a) and b) axes. The fragment detectors are placed
symmetrically with respect to the beam axis and their centers are located at
angles $\theta_{\rm lab.}$~$\simeq \pm $~45$^{o}$, then the c) axis can be
approximatively considered as the molecular axis parallel to the relative
vector between the two centers of the out-going binary fragments. In
Fig.8 the results of the $\gamma$-ray angular correlations for the
mutual excitation exit-channel are shown. It is interesting to note that almost
identical results have been also obtained for the single excitation
exit-channel. The minima observed in a) and b) at 90$^{o}$ imply that the
intrinsic spin vectors of the 2$^{+}$ states lie in the reaction plane and are
perpendicular to the orbital angular momentum. So the value of the angular
momentum remains close to L = 38 ${\rm \hbar}$ for the two exit channels, in
agreement with Fig.7. The maximum around 90$^{o}$ in c) suggests
that the $^{28}$Si spin vectors are parallel to the fragment directions with
opposite directions. Such disalignment of the fragment spins are, of course,
not usual in deep inelastic processes [12], but a very long life-time of the
resonance might allow large microscopic fluctuations. In the present
experiment the feeding of the $^{28}$Si states are also measured. The study of
the feeding of the bands of $^{28}$Si, has revealed that the $^{28}$Si is
dominated by {\it oblate} deformation [9,10,11]. 
\par
Recently A.H. Wuosmaa et al. [13] have observed that the prolate-prolate system
$^{24}$Mg~$+$~$^{24}$Mg is characterised by spin alignment and this is in
contrast with our results in the oblate-oblate system $^{28}$Si~+~$^{28}$Si. In
the following section we will attempt to explain spin disalignment in the
oblate-oblate system $^{28}$Si~+~$^{28}$Si and spin alignment in the
prolate-prolate system $^{24}$Mg~$+$~$^{24}$Mg by using a new molecular model. 
\vskip 0.5cm
\centerline{\Large \sl \bf C. Molecular Model }
\centerline{\large \sl ``Interpretation of Spin Disalignment : Butterfly Motion''}
\vskip 0.5cm
Quasi-molecular resonances in heavy-ion scattering of $^{28}$Si~+~$^{28}$Si and
$^{24}$Mg~$+$~$^{24}$Mg are examined in the framework of a new molecular model.
The general ideas of the molecular model are based on the following
considerations : the occurence of elongated but stable dinuclear configurations
and their characteristic normal-mode of motions at equilibrium is proposed to
be the origin of the observed resonant structures. The problem is ''how to find
and how to describe these motions ?''. For this purpose, we define a rotating
molecular frame of the dinuclear system, and consider collective motions of the
intrinsic states [9,14,15,16,17]. 
\par
A stable configuration of the $^{24}$Mg~$+$~$^{24}$Mg scattering is found to be
a {\it pole-pole} (P-P) configuration, due to the prolate shape of the
$^{24}$Mg interacting nuclei. Around the equilibrium various dynamical modes
which may appear could be responsible for the observed narrow resonances [13].
In an oblate-oblate system $^{28}$Si~+~$^{28}$Si (the $^{28}$Si nucleus has an
oblate shape [9,10,11]) with high spins such as 38${\rm \hbar}$, an equator-equator
(E-E) touching configuration is favored. Hence it is interesting to investigate
molecular modes around such a new type of equilibrium. In Fig.10, an
oblate-oblate configuration is displayed, where we have introduced a rotating
molecular frame, the $z^{'}$-axis of which is parallel to the relative vector
of the two interacting nuclei. For the sake of simplicity, we consider the
system of two identical deformed nuclei with a constant deformation and axial
symmetry, and then we have seven degrees of freedom. Three of them are from the
relative vector {\trpolice R} = ($R,\theta_{1},\theta_{2}$). Collective degrees
of freedom of deformed nuclei are the orientations of the symmetry axes, which
are described by the Euler angles ($\alpha_{1}, \beta_{1}$) and
($\alpha_{2},\beta_{2}$). $\alpha_{1}$ and $\alpha_{2}$ are combined into
$\theta_{3}$~=~($\alpha_{1} + \alpha_{2}$)/2 and $\alpha$ = ($\alpha_{1} -
\alpha_{2}$)/2. Thus we have ($q_{i}$) =
($\theta_{1},\theta_{2},\theta_{3}$,$R$,$\alpha,\beta_{1},\beta_{2}$), where
$\theta_{i}$'s are the Euler angles of the molecular frame with four other
internal variables. Consistent with the coordinate system, we introduce
rotation-vibration type wave functions as the basis, 
\begin{equation}
\Psi \sim D^{J}_{MK}(\theta_{i})\chi_{K}(R,\alpha,\beta_{1},\beta_{2}).
\end{equation}
By solving the internal motions for $\chi_{K}(R,\alpha,\beta_{1},\beta_{2})$,
{\it butterfly and anti-butterfly} vibrational modes, and $K$-and
twisting-rotational modes (associated variables $\theta_{3}$ and $\alpha$, and
quantum numbers $K$ and $\nu$, respectively) are obtained (Details are given in
Ref. [9,16,17]). Fig.11 and 12 display recent theoretical analyses for
$^{28}$Si~+~$^{28}$Si and $^{24}$Mg~$+$~$^{24}$Mg in the excited states
(2$_{1}^{+}$,2$_{1}^{+}$) for the normal-mode motions by the molecular model. 
\par
\noindent{\trpolice  For $^{28}$Si~+~$^{28}$Si :}  
{
Fig.11 shows (see panel B) that there is an extreme of the
probabilities for channel spin of I = 0 for Butterfly motion. This is in good
agreement with the experimental angular distribution with L = J in
Fig.7. Furthermore, very recently, the molecular model is developed
[14] to include the K-mixing, and a tilting mode is obtained. The results
exhibit strong concentration in "m = 0" states [14], which may be in good
agreement with the experimental $\gamma$-ray distribution. Work is in progress
for a reasonable comparison between the data and the calculated results. It
seems anyway that in the $^{28}$Si~+~$^{28}$Si system (oblate-oblate), the spin
disalignment could result from a Butterfly motion and tilting mode [9,10,11]. }
\par
\noindent{\trpolice  For $^{24}$Mg~$+$~$^{24}$Mg : } 
Fig.12 shows that for the $^{24}$Mg~$+$~$^{24}$Mg system
(prolate-prolate) the different vibrational modes result in channel spin
probabilities which are more equal. In this case the Butterfly motion is not
anymore the dominant mode.
\par 
\noindent As a consequence the qualitative differences between
$^{28}$Si~+~$^{28}$Si and $^{24}$Mg~$+$~$^{24}$Mg in the (I$_{1}$,I$_{2}$) =
(2$_{1}^{+}$,2$_{1}^{+}$) channel are clearly obtained. 
\vskip 0.3cm
\centerline{\Large \sl \bf IV. Discussion and Comparison between Different Systems}
\centerline{\large \sl ``\ $^{12}$C~$+$~$^{12}$C, 
$^{24}$Mg~$+$~$^{24}$Mg and $^{28}$Si~$+$~$^{28}$Si\ ''}
\vskip 0.3cm
The first observation of the correlated spin orientation (alignment spin) has
been in $^{12}$C~$+$~$^{12}$C quasi-molecular resonances in 1985 [18]. In this case
the correlation between the spin orientations of the two $^{\rm 12}$C(2$^{\rm
+}$) nuclei in mutual inelastic $^{12}$C~$+$~$^{12}$C scattering has been
deduced from the directional correlations of the particle-coincident
$\gamma$-ray measured with the Heidelberg crystal-ball detector. Resonances in the cross
section are found to be nearly uniquely associated with the mutually aligned
component. The strong resonance structure as well as the characteristic
behavior of the angular distributions of this reaction component are suggestive
of the formation of a rotating complex in the sticking configuration. Because
of the deformed shape of $^{12}$C (oblate), the sticking condition, which is a
well-known classical characteristic of deep-inelastic collisions in heavier
systems [19], becomes relevant for the quantum mechanical phenomenon of
$^{12}$C~$+$~$^{12}$C resonances. 
\par
\hspace*{0.25cm} Two years after the observation of the spin alignment in 
$^{12}$C~$+$~$^{12}$C molecular resonances, A.~Wuosmaa and collaborators [13]
have measured the single and correlated magnetic-substate population parameters
for 2$^{+}$ and 2$^{+}$-2$^{+}$ excitations in $^{24}$Mg~$+$~$^{24}$Mg
(prolate-prolate system) in the region of two strong resonances observed in
inelastic scattering. They have performed their $\gamma$-ray angular
correlation and angular-distribution measurements using the Oak Ridge Spin
Spectrometer. This device is a multidetector NaI spectrometer comprised of 70
NaI $\gamma$-ray detectors. These data have provided spectroscopic information
relatively uncontaminated by nonresonant amplitudes, and allow spin assignment
of J$^{\pi}$ = 36$^{+}$ for two resonances E$_{\rm c.m.}$~=~45.70 and 46.65
MeV. The angular correlation data for the mutual 2$^{+}$ inelastic scattering
channel suggest a dominant decay ${\rm \ell}$ value of ${\rm \ell}$~=~34~${\rm
\hbar}$ for both resonances. Correlated spin alignment data for this channel
confirm the expectations for the relationship between angular momentum coupling
and spin alignment for these resonances. The relatively high spin values
suggest a resonance configuration in which the two $^{24}$Mg nuclei interact
pole-to-pole, allowing the system to sustain a large amount of angular
momentum. This results is well corroborated by the molecular model~[17]. 
\par
In the present experiment $^{28}$Si~$+$~$^{28}$Si (oblate-oblate system),
performed at the resonance energy E$_{\rm c.m.}$~= 55.8~MeV high-resolution
fragment-fragment-$\gamma$ data have been collected with the Eurogam Phase~II
multi-detector array. It is to our knowledge the first time that the
disalignment spin has been observed in a heavy ion reaction. This has been
primarly shown in the measured angular distributions of the elastic, inelastic
2$^{+}$, and mutual excitation channels 2$^{+}$-2$^{+}$, which are dominated by
a unique and pure partial wave with L = 38\ ${\rm \hbar}$, and has been
confirmed by measuring their particle-$\gamma$ angular correlations with
Eurogam Phase~II. Within the molecular model the spin disalignment is taken to
suggest a dominance of the Butterfly mode in the vibrational motion of the
observed resonance. Therefore a stable configuration is infered to be an
elongated one, namely an equator-to-equator touching configuration. 
\par
The comparison between the three symmetric systems $^{12}$C~$+$~$^{12}$C,
$^{24}$Mg~$+$~$^{24}$Mg and $^{28}$Si~$+$~$^{28}$Si shows an interesting
contrast in the spin orientation at resonance energies. The results indicate
that $^{28}$Si~$+$~$^{28}$Si (oblate-oblate system) is characterised by spin
disalignment in contrast to the observed spin alignment for
$^{12}$C~$+$~$^{12}$C system (oblate-oblate) and $^{24}$Mg~$+$~$^{24}$Mg system
(prolate-prolate). The molecular model calculations explain the spin
disalignment in the oblate-oblate $^{28}$Si~$+$~$^{28}$Si system by a Butterfly
motion. Now the question arises, ''why is the orientation of spin in the two
oblate-oblate systems $^{28}$Si~$+$~$^{28}$Si and $^{12}$C~$+$~$^{12}$C
different~?''. This problem would be addressed both experimentally and
theoretically in a near future. 

\vskip 0.7cm
\noindent{\large \bf Acknowledgments} : We would like to acknowledge the
VIVITRON operators for providing us with well focussed $^{28}$Si beams. Two of
us (C.B. and R.N.) would like to acknowledge Professor R.R.~Betts and Professor
~W.~Von~Oertzen for useful discussions on various aspects of this work. 
\newpage

\centerline{ \large \bf REFERENCES}
\vskip 0.5cm
\normalsize
\hspace*{-1.15cm} [1] R.R. Betts et al., Phys. Rev. Lett.  {\bf 47}, 23(1981).
\newline
\hspace*{-0.5cm} [2] R.W. Zurm\"{u}hle et al., Phys. Lett. {\bf B129}, 384(1983).
\newline
\hspace*{-0.5cm} [3] P.H. Kutt et al., Phys. Lett {\bf B155}, 384(1983).
\newline
\hspace*{-0.5cm} [4] A.H. Wuosmaa et al., Phys. Rev. {\bf C36}, 1011(1987).
\newline
\hspace*{-0.5cm} [5] G. Leander et al., Nucl. Phys. {\bf A239}, 93(1975),
\newline
\hspace*{0.15cm}
M.E. Faber et al., Phys. Scr. {\bf 24}, 189(1981).
\newline
\hspace*{-0.5cm} [6] R.R. Betts et al., Proc. Inter. Conf. on Nuclear 
Physics, 
\newline
\hspace*{0.15cm}
Stony Brook, New York, 347(1983).
\newline
\hspace*{-0.5cm} [7] R.R. Betts et al., Phys. Lett. {\bf B100}, 177(1981).
\newline
\hspace*{-0.5cm} [8] R.R. Betts et al., Phys. Rev. Lett. {\bf 46}, 313(1981).
\newline
\hspace*{-0.50cm} [9] R. Nouicer, Ph.D. Thesis, Strasbourg University, 
Report IReS 97-35. 
\newline
\hspace*{-0.70cm} [10] R. Nouicer, C. Beck et al., Proc. of Int. Scho. Semi. on Heavy Ion 
Physics, 
\newline
\hspace*{0.15cm}
Dubna, Russia 1997, to be published, Report IReS 97-32, {\bf nucl-ex/9711002} (1997)   
\newline
\hspace*{-0.70cm} [11] C. Beck, R. Nouicer et al., Report IReS 97-22, {\bf nucl-ex/9708004} (1997)
\newline
\hspace*{-0.70cm} [12] J. Randrup, Nucl. Phys. {\bf A447}, 133(1985).
\newline
\hspace*{-0.7cm} [13] A.H. Wuosmaa et al. Phys. Rev. Lett. {\bf 58}, 1312(1987).
\newline
\hspace*{-0.7cm} [14] Y. Abe and E. Uegaki, private communications (1998).
\newline
\hspace*{-0.7cm} [15] E. Uegaki and Y. Abe, Phys. Lett. {\bf B340}, 143(1994).
\newline
\hspace*{-0.7cm} [16] E. Uegaki and Y. Abe, Prog. Theor. Phys. {\bf 90}, 615(1993).
\newline
\hspace*{-0.7cm} [17] E. Uegaki and Y. Abe, Phys. Lett. {\bf B231}, 28(1989)
\newline
\hspace*{-0.7cm} [18] D. Konnerth et al., Phys. Rev. Lett. {\bf 55}, 588(1985).
\newline
\hspace*{-0.7cm} [19] A. Gobbi and N\"orenberg, in Heavy Ion Collisions, 
\newline \hspace*{0.15cm}Vol. 2, edited by R. Bock (North-Holland, Amsterdam, 1980), p128.
\newpage

\centerline{\bf FIGURE CAPTIONS}

\bigskip

Fig.1 : Excitation functions data for $^{28}$Si + $^{28}$Si scattering averaged
over 67$^{o} \le \theta_{\rm c.m.} \le $ 95$^{o}$ and for $^{24}$Mg + $^{24}$Mg
scattering averaged over 66$^{o} \le \theta_{\rm c.m.} \le $ 93$^{o}$. 

Fig.2 : Potential energy surface for $^{56}$Ni and for $^{48}$Cr calculated in
the Nilson-Strutinsky prescription. The $\beta$ is the deformation parameter
and the M$_{\rm r}$ = M/56 (or M/48) is the mass-asymmetry where M is the mass
of the fragment in the exit-channel. 

Fig.3 : Detection system used in the experiment $^{28}$Si $+$ $^{28}$Si. \\
A) General view of Eurogam Phase II multi-detector array (gamma-ray detectors) \\
B) inside view of the reaction chamber where four position-sensitive detectors
(fragment detectors) can be observed. 

Fig.4 : Two-dimensional energy spectrum E$_{4}$ versus E$_{3}$ of the fragments
detected by one pair of position-sensitive Si detectors for the
$^{28}$Si~$+$~$^{28}$Si reaction. 

Fig.5 : Mass spectra corresponding to the region 1. \\
A) corresponds to the reaction products of $^{28}$Si~$+$~$^{28}$Si. \\  
B) products arising from of the reaction $^{28}$Si on the 
contamination $^{16}$O. \\
The peaks have been identified by using $\gamma$-ray spectra. 

Fig.6 : Ejectile energy E$_{3}$ versus excitation energy E$_{X}$
two-dimensional plot of the $^{28}$Si~$+$~$^{28}$Si symmetric exit-channel.. 

Fig.7 : Experimental angular distributions of the elastic (E$_{X} \approx$  0
MeV), inelastic (E$_{X} \approx$  1.7~MeV) and mutual excitations (E$_{X}
\approx$  4, 6 and 10 MeV) for the $^{28}$Si~$+$~$^{28}$Si symmetric
exit-channel. The solid curves represent pure squared Legendre polynomial
$[$P$_{L}(\cos \theta_{c.m.} )]^{2}$ with L = 38. 

Fig.8 : Experimental $\gamma$-ray angular correlations of the
(2$^{+}_{1}$,~2$^{+}_{1}$) states in the resonance region of the $^{28}$Si +
$^{28}$Si exit channel for three quantization axes. 

Fig.9 : Diagram illustrating the quantification axes $\vec{oa}$ ,$\vec{ob}$ and
$\vec{oc}$. The PSD 3 and 4 in coincidence are represented in the reaction
plane. The spins directions are indicated according to experimental results as
shown in Fig.8. 

Fig.10 : Dinuclear configuration (oblate-oblate system) and the coordinates in
the rotating molecular frame [9,15]. 

Fig.11 : Molecular model prediction for probability distributions of the
$^{28}$Si~$+$~$^{28}$Si ([ 2$^{+}_{1} \otimes $ 2$^{+}_{1}$] with L = J - I)
system versus channel spin I. 

Fig.12 : Molecular model prediction for Probability distributions of the
$^{24}$Mg~$+$~$^{24}$Mg ([ 2$^{+}_{1} \otimes $ 2$^{+}_{1}$] with L~=~J~-~I)
system versus channel spin I.

\end{document}